\documentclass[twocolumn,showpacs,preprintnumbers,amsmath,amssymb]{revtex4}

\usepackage{graphicx}
\usepackage{dcolumn}
\usepackage{bm}

\newcommand{\ua}{\uparrow}
\newcommand{\da}{\downarrow}

\begin{document}


\title{The many-electron ground-state determines uniquely the potential in \\
Spin-Density-Functional Theory for non-collinear magnetism}

\author{Nikitas I. Gidopoulos}
 \email{n.gidopoulos@rl.ac.uk}
\affiliation{%
ISIS Facility, Rutherford Appleton Laboratory, 
Chilton, Didcot, Oxon, OX11 0QX, England, U.K.
}%

\date{\today}

\begin{abstract}
Since Spin Density Functional Theory was first proposed, but also recently, examples were constructed to show that 
a spin-potential may share its ground state with other spin-potentials.
In fact, 
for collinear magnetic fields and systems with fixed magnetization,
the mapping between potentials and ground states is 
invertible, provided the magnetization is not saturated and that spin-potentials are determined within a spin-constant. 
We complete the proof that the mapping is invertible also for non-collinear magnetic fields and systems 
with more than one electron. 
We then discuss the non-collinear exchange and correlation energy functional and suggest improvements.
\end{abstract}

\pacs{71.15.Mb, 31.15.Ew}
\maketitle

The remarkable success of Density Functional Theory (DFT) \cite{hk,ks,mel,dftrev}, as a tool for electronic 
structure calculations in solid state physics and quantum chemistry, is owed to its efficiency but also to 
the soundness of the underlying principles. 
In fact the quest to refine the foundations of the theory often results naturally in better understanding and 
improving the accuracy of the approximation. 
The aim of this work is to present such an example.
 
So far, it appeared the sound basis of ground state (gs) DFT is not shared by gs Spin Density Functional Theory (SDFT) \cite{ks,vonbarth,rajagopal1,perdew,kubler,sandratskii}, commonly 
used when the electronic system of interest lies in a weak external magnetic field, 
or even when there is no external magnetic field but the electrons in the system do not form closed shells. 
Although formally SDFT is a straightforward extension of the original theory, 
the analogue of DFT's Hohenberg Kohn (HK) theorem \cite{hk}, 
establishing the 1-1 correspondence between the set of spin-potentials, 
$\{ {\mathcal V} \}$, and the set of gs spin-densities, 
$\{ \varrho \}$, has not been established yet for non-collinear magnetic fields. 
By the spin-potential we mean the potential and magnetic field, ${\mathcal V} = 
( V ; {\bf B} )$, while 
the spin-density represents the charge and magnetization densities 
$\varrho = ( \rho; {\bf m} )$. 

Traditionally, invertibility of the mapping between $\{ {\mathcal V} \}$ and $\{ \varrho \}$ 
is proven in two steps. 
Firstly showing that different spin-potentials have different ground states, i.e., 
the mapping between the set of spin-potentials $\{ {\mathcal V }  \}$ 
and the set of ground states $\{ \Psi \}$ is invertible.
Secondly, showing that different ground states, arising from different spin-potentials, 
have different densities, i.e., that the mapping between 
$\{ \Psi \}$ and $\{ \varrho \}$ is also 1-1.   

It is straightforward to show that the second mapping is invertible. 
In the publication where von Barth and Hedin \cite{vonbarth} proposed SDFT, they also constructed, for the single-electron gs $\Psi$ of any spin-potential, a whole class of different (non-collinear) spin-potentials that admitted $\Psi$ as an eigenstate, and under rather mild conditions as their gs (see however \cite{rajagopal2}), concluding that the first step in the proof of the HK theorem could not hold in general. 
Recently, interest revived on the same issue \cite{capelle,eschrig,argaman,mydsdft,kohnsavinullrich,ullrich}.
In the case of collinear magnetic fields, Capelle and Vignale \cite{capelle} gave several examples 
where the spin-potential was not determined uniquely by the spin-density. 
Eschrig and Pickett \cite{eschrig} studied collinear and non-collinear magnetic fields.
They distinguished two cases, when the gs is a pure-spin (p-s) state or an impure-spin (imp-s) state. Pure-spin states are states which through a local rotation of the spin-coordinates may be transformed to have a definite number of spin-up and down electrons. 
Impure-spin states are states which cannot be transformed to have a definite number of spin-up and down electrons under any local spin-rotation.
It was found that when the gs of the Hamiltonian ${\hat {\mathcal H}}$ is an imp-s state, there is no ambiguity (apart from an overall constant) in determining the spin-potential in ${\hat {\mathcal H}}$ by the ground state. 
However they also discovered p-s states are eigenstates of an operator which, in the laboratory unrotated spin-space, corresponds to a non-collinear magnetic field  with constant magnitude and concluded the spin-potential in ${\hat {\mathcal H}}$ cannot be determined uniquely by the ground state. 

If these findings did hold in general, the meaningfulness of the Kohn Sham (KS) scheme would be questioned, since in the latter, the non-interacting KS spin-potential is not known and must be determined together with the KS state that yields the gs spin-density. 

Recent work gives hope that the mapping is invertible. Numerical investigations by Ullrich \cite{ullrich} on lattices give evidence that in the continuum limit and for non-collinear magnetic fields, non-uniqueness is very rare if not completely absent. Further, the study of finite temperature SDFT by Argaman and Makov \cite{argaman}, in the limit of vanishing temperature (provided the limit is not singular) shows that the mapping between ${\mathcal V} $ and ${\varrho}$ is invertible. 

In the first part of the paper we resolve the issue proving conclusively that the mapping is indeed invertible.

There is no ambiguity regarding the invertibility of the mapping between potentials and 
ground states in SDFT for collinear fields. 
It should be noted beforehand that for collinear magnetic fields the eigenstates of ${\hat {\mathcal H}}$ are p-s states, since the spin-up and spin-down particle-number operators, ${\hat N}_{\ua}$, ${\hat N}_{\da}$, independently, commute with ${\hat {\mathcal H}}$. Capelle and Vignale \cite{capelle} discovered two cases of freedom in determining the spin-potential:
 
First, as in DFT where the trivial density, $\rho({\bf r}) = 0$, cannot determine the potential, $V({\bf r})$, 
in the same way here a trivial component of the spin-density cannot determine the corresponding 
component of the spin-potential, i.e., for a system of electrons that are all 
spin-polarized in one direction, say, $\rho_{\ua }({\bf r}) > 0$, $\rho_{\da }({\bf r}) = 0$,  
the spin-component of the potential in the other direction, $V_\da({\bf r})$, is completely 
arbitrary (as long as the system remains fully polarized). 
This kind of indeterminacy cannot have any practical consequence as there are no electrons to experience 
the arbitrariness.

Second, together with the freedom of a constant shift
in the potential (because ${\hat N}_{\ua} + {\hat N}_{\da}$ 
commutes with ${\hat {\mathcal H}}$), there is an additional 
freedom of a constant shift in the magnetic field, $ B_0$  
appearing because the magnetization operator ${\hat N}_{\ua} - {\hat N}_{\da}$, 
commutes with ${\hat {\mathcal H}}$.
According to Capelle and Vignale, this is an example of a systematic non-uniqueness arising from a conserved quantity (as opposed to an accidental non-uniqueness like the example of the perfectly spin-polarized system.) 
The formulation of SDFT can be corrected to respect this freedom by
constraining, at the functional differentiations, both the charge and the magnetization densities to integrate separately to fixed number of particles \cite{mydsdft}.
Different KS calculations must be performed for systems with fixed magnetizations $N^\ua - N^\da$ (fixed spin moment calculations), and the gs is given by the system with the lowest KS total energy \cite{mydsdft}.

Still, the challenging question in collinear SDFT, in order to establish the $1-1$ mapping 
is whether this is all the freedom available, or whether there could be other kinds of non-uniqueness present. 
It was shown that for continuous spin-potentials this does not happen \cite{eschrig,mydsdft}. 

The general case of non-collinear SDFT is not very different. 
We focus first on von Barth and Hedin's example.
For any single-electron state $\Psi$ they constructed a single-particle spin-potential ${\mathcal V}'$ which acting on $\Psi$ gives zero, 
\begin{equation} \label{vbh}
{\mathcal V}' \, \Psi = 0.
\end{equation}
Hence, if $\Psi$ is the ground  state of a spin-Hamiltonian ${\hat {\mathcal H}}$, then $\Psi$ will be an eigenstate of 
${\hat {\mathcal H}} + \lambda \, {\mathcal V}'$. A whole class of ${\mathcal V}'$ was found and it was argued that small enough $\lambda$ must exist for which $\Psi$ is the gs of ${\hat {\mathcal H}} + \lambda \, {\mathcal V}'$.    
Von Barth and Hedin did not explain how they constructed ${\mathcal V}'$, but it is easy to do so:
any single-particle state $\Psi$ with spinor wave-function 
$\big(\,^{{ \psi}_{\ua}({\bf r})}_{ { \psi}_{\ua}({\bf r})} \big)$ 
defines its own (local) orientation in spin-space, call it spin-up,
$| \!\! \ua \rangle$ = $\big(\,^{{ \psi}_{\ua}({\bf r})}_{ { \psi}_{\da}({\bf r})} \big)$
/$\sqrt{|{ \psi}_{\ua}({\bf r})|^2 + |{ \psi}_{\da}({\bf r})|^2}$. 
The spin-down orientation is also defined: 
$| \!\! \da \rangle=$ $\big(\,^{\ { \psi}_{\da}
^*({\bf r})}_{ -{ \psi}_{\ua}^*({\bf r})} \big)  /\sqrt{|{ \psi}_{\ua}({\bf r})|^2 + 
|{ \psi}_{\da}({\bf r})|^2 }$. 
Then, $\Psi$ is an eigenstate of the single-particle spin-down number operator $| \!\! \da \rangle \langle \da \! |$ with eigenvalue zero and for any $ V'({\bf r}) $
the spin-potential $ {\mathcal V}' = V'({\bf r}) \, | \!\! \da \rangle \langle \da \! |$ maps $\Psi$ to zero. 
It is evident this example is a generalisation to the non-collinear case 
of the accidental non-uniqueness for systems fully saturated in one spin-direction, 
leaving the potential in the other spin-direction undetermined \cite{argaman}.
To extend von Barth and Hedin's example to many-electron systems, 
one would need to construct fully spin-polarized p-s states and therefore to obtain  a 
local rotation in spin-space that reduces ${\hat {\mathcal H}}$ to spin-diagonal form.

We point out that the discovery by Eschrig and Pickett relies exactly on the 
assumption that this is possible. 
In fact criticism on the reasoning by Eschrig and Pickett (Footnote 18 in \cite{argaman}) misses the point, as it is restricted to 
Hamiltonians that cannot be transformed to spin-diagonal form. 

So, given the proof \cite{eschrig} that imp-s states do not lead to any kind of non-uniqueness, 
the question about the invertibility of the mapping between spin-potentials and 
ground-states for non-collinear spin Hamiltonians boils down to the existence 
or not of a local transformation that spin-diagonalizes 
${\hat {\mathcal H}}$.   
We consequently prove that a Hamiltonian, 
\begin{equation} \label{one}
{\hat {\mathcal H}} = {\hat T} + \lambda \, {\hat V}^{\rm ee} + {\hat V} + {\hat  B} \, ,
\end{equation}
cannot be transformed to spin-diagonal form, unless the magnetic field is collinear.

${\hat T}$, ${\hat V}^{\rm ee}$ in (\ref{one}) 
are the kinetic energy and electron repulsion 
operators, $\lambda =1 $ ( 0 ) for an interacting (non-interacting) system and
${\hat V}$, ${\hat B}$ are the operators for the potential and the magnetic field.
We shall focus on
\begin{equation} \label{T}
{\hat T}  
  = 
- { \hbar^2 \over 2 m } \sum_{\tau = \ua , \da} \int \!\! d{\bf r} \, 
 {\hat \psi}_{\tau}^{\dagger} ({\bf r})
\, \nabla^2
	{\hat \psi}_{\tau} ({\bf r}) 
\end{equation}
and
\begin{equation} \label{B}
{\hat B}
  =
- \int \! d{\bf r} \, {\hat {\bf m}} ({\bf r}) \cdot {\bf B} ({\bf r}) 
\end{equation}
where, ${\hat \psi}^{\dag}_{\tau }({\bf r}), {\hat \psi}_{\tau }({\bf r})$ are 
creation and annihilation fermion field operators, 
${\hat {\bf m}} ({\bf r})$ is the magnetization density operator
\begin{equation} 
{\hat {\bf m}} ({\bf r})
  =
 - \mu_0 \, 
\left( {\hat \psi}_{\ua}^{\dagger}({\bf r}) \, {\hat \psi}_{\da}^{\dagger}({\bf r}) \right) 
\, 
\mbox{\boldmath $\sigma$}
\,
\left( 
	\!\! \begin{array}{c}
	{\hat \psi}_{\ua}({\bf r}) \\
	{\hat \psi}_{\da}({\bf r})
	\end{array} \!\!
\right) 
\end{equation}
 and 
{\boldmath $\sigma$}  is the vector of 
$2 \times 2$ Pauli spin-matrices,  \mbox{\boldmath $\sigma$}  = $(\sigma_x, \sigma_y, \sigma_z) $.

We introduce the general unitary transformation that rotates locally at every point in real space the spin degrees of freedom 
\begin{equation} \label{U}
{\sf U} ({\bf r}) = 
\left( 
	\begin{array}{cc}
	e^{i \theta ({\bf r})} \cos{\omega({\bf r})} & - e^{i \phi ({\bf r})} \sin{\omega({\bf r})} \\
	e^{- i \phi ({\bf r})} \sin{\omega({\bf r})} & e^{- i \theta ({\bf r})} \cos{\omega({\bf r})}   
	\end{array}
\right)
\end{equation}
$\theta({\bf r})$, $\phi({\bf r})$, $\omega({\bf r})$ are real functions. 
We have 
$
{\sf U} ({\bf r}) \, {\sf U}^{\dag} ({\bf r}) 
$ = 
${\sf U}^{\dag} ({\bf r}) \, {\sf U} ({\bf r}) = {\sf I} 
$. 
Rotated in spin-space second quantized fermion field operators can be defined: 
\begin{eqnarray} 
\left( 
{\hat \Psi}_{\ua}^{\dagger}({\bf r}) \, {\hat \Psi}_{\da}^{\dagger}({\bf r}) 
\right) &
= &
\left( 
{\hat \psi}_{\ua}^{\dagger}({\bf r}) \, 
{\hat \psi}_{\da}^{\dagger}({\bf r}) 
\right) 
{\sf U}^{\dag} ({\bf r}) \, ,  \label{rot1}
 \\  
\left( 
	\! \begin{array}{c}
	{\hat \Psi}_{\ua}({\bf r}) \\
	{\hat \Psi}_{\da}({\bf r})
	\end{array} \!
\right) & 
= &
{\sf U} ({\bf r})
\left( 
	\! \begin{array}{c}
	{\hat \psi}_{\ua}({\bf r}) \\
	{\hat \psi}_{\da}({\bf r})
	\end{array} \!
\right) \label{rot2}
\end{eqnarray}

Is there a ${\sf U}$ that diagonalizes ${\hat {\mathcal H}}$ in spin-space?

A rotation in spin-space may transform locally the magnetic field operator ${\hat B}$ to diagonal form.
${\hat V}$ and ${\hat V}^{\rm ee}$ are independent of spin and do not change form.
However, the kinetic energy operator $\hat T$ in (\ref{T}) is not invariant under the gauge-transformation 
(\ref{U}) and the transformed operator represents the motion of a particle 
in a gauge vector potential $\mbox{\boldmath ${\sf A}$} ({\bf r})$ in spin-space
(omit obvious $\bf r$ dependence):
\begin{equation} \label{Tgauge}
{\hat T} =   {\hbar^2 \over 2 m} \int d{\bf r} 
\left( {\hat \Psi}_{\ua}^{\dagger} 
\, {\hat \Psi}_{\da}^{\dagger} 
\right) 
\left[ -i \mbox{\boldmath $\nabla$} + \mbox{\boldmath ${\sf A}$} 
\right]^2  
\left( 
	\!\!\! \begin{array}{c}
	{\hat \Psi}_{\ua} 
	\\
	{\hat \Psi}_{\da} 
	\end{array} \!\!\!
\right)  ,  
\end{equation}
where 
$
\mbox{\boldmath ${\sf A}$} ({\bf r}) = - i  \, 
{\sf U} ({\bf r})
\mbox{\boldmath $\nabla$}
{\sf U}^{\dag} ({\bf r}) .
$ 

We further consider the rotated spin-up, down number operators 
$
{{\hat N}_{\sf U}}\!_{\, \tau}
\doteq \int d {\bf r} \, {\hat \Psi}_{\tau}^{\dagger}({\bf r}) {\hat \Psi}_{\tau}({\bf r})
$, $\tau = \ua , \da$,   
and the magnetization operator in rotated spin-space: 
\begin{equation}
{\hat b} = {{\hat N}_{\sf U}}\!_{\, \ua} - {{\hat N}_{\sf U}}\!_{\, \da} \, .
\end{equation}
Rotating back to the original spin space we see that $\hat b$ is 
the operator of a magnetic field ${\bf b}({\bf r})/\mu_0$ (\ref{B}):   
\begin{equation}
{\hat b} = - \int d{\bf r} \, {\hat {\bf m}} ({\bf r}) \cdot { {\bf b} ({\bf r}) \over \mu_0 } 
\end{equation} 
${\bf b} ({\bf r}) $ is in general non-collinear,  
%
$b_x({\bf r})  =  - \sin{2 \omega({\bf r})} $ $   \cos{\big( \theta({\bf r})  - \phi({\bf r}) \big)} $, 
$ b_y({\bf r})  =  - \sin{2 \omega({\bf r})} \,  \sin{ \big( \theta({\bf r}) - \phi({\bf r}) \big)} $, 
$ b_z({\bf r})  =  \cos{2 \omega({\bf r})} $,
and has unit magnitude $b^2({\bf r}) = 1$.  

We obtain tautologically the argument of Eschrig and Pickett. A p-s state (in rotated spin-space) is by definition an eigenstate of the rotated magnetization ${\hat b}$, which in the unrotated 
spin-space represents a non-collinear magnetic field with constant magnitude.
Of course, the question is whether p-s states exist, or equivalently whether an operator $\hat b$ exists that 
commutes with ${\hat {\mathcal H}}$ \cite{path}. In that case,  ${\hat {\mathcal H}}$ could be brought to spin-diagonal form.  
Indeed in Ref. \cite{eschrig}, it is explicitly assumed that such $\hat b$ exists ($\hat b$ is denoted ${\hat U}_o$, defined by Eq. 15 in \cite{eschrig}). 

%
To answer the question we need the commutator $[ {\hat {\mathcal H}} , {\hat b}]$. We have $[{\hat V}^{\rm ee} , {\hat b}] 
= [{\hat V} , {\hat b}] = 0$ and
\begin{equation} \label{TU} 
[ {\hat T} , {\hat b}]
 = -i \hbar \sum_{\alpha = x,y,z} \int d{\bf r}   \ 
{\hat {\bf j}}_{\alpha} ({\bf r}) \cdot \mbox{\boldmath $\nabla$} b_{\alpha}({\bf r}) 
\end{equation}
\begin{equation} \label{BU}
[{\hat B} , {\hat b}] =  
- 4 i \int d{\bf r} \ {\hat {\bf m}} ({\bf r}) \cdot    
{ {\bf B}({\bf r}) \times {\bf b}({\bf r}) \over \mu_0 }
\end{equation}
where, 
\begin{eqnarray}
&&
 {\hat {\bf j}}_{\alpha }  ({\bf r})
 =  {i  \hbar \over 2 m}  
\left\{
\left[ \mbox{\boldmath $\nabla$}
\left( {\hat \psi}_{\ua}^{\dagger} ({\bf r})
\, 
{\hat \psi}_{\da}^{\dagger} ({\bf r})
\right) \right]
\sigma_{\alpha }
\left( 
	\!\!\! \begin{array}{c}
	{\hat \psi}_{\ua}
	({\bf r}) 
	\\
	{\hat \psi}_{\da}
	({\bf r})
	\end{array} \!\!\!
\right) 
\right.  \\
&& \label{j}
\ \ \ 
\left.
 - 
\left( {\hat \psi}_{\ua}^{\dagger}
 ({\bf r}) \, 
{\hat \psi}_{\da}^{\dagger}
({\bf r}) 
\right) 
\sigma_{\alpha }
\left[
\mbox{\boldmath $\nabla$}
\left( 
	\!\!\! \begin{array}{c}
	{\hat \psi}_{\ua}
	({\bf r}) 
	\\
	{\hat \psi}_{\da}
	({\bf r})
	\end{array} \!\!\!
\right)
\right]
\right\} \, , \ \alpha = x,y,z \nonumber
\end{eqnarray}

The two commutators correspond to operators for different physical quantities, $i \, [ {\hat T} , {\hat b}]$ describes the interaction of currents with vector potentials and $i \, [{\hat B} , {\hat b}]$ the energy of a magnetic moment in a magnetic field and hence they cannot cancel each other. In order that $[ {\hat {\mathcal H}} , {\hat b}] = 0$,  
each commutator must vanish separately. 
From (\ref{TU}) we have
$\mbox{\boldmath $\nabla$} b_{x, y, z}({\bf r}) = 0$,           
i.e., ${\bf b}$ is independent of 
$\bf r$. 
From (\ref{BU}) we have  
${\bf B}({\bf r}) \times {\bf b} = 0$ and 
${\bf B}({\bf r})$ must be collinear, as it has to be parallel to ${\bf b}$ which does not depend on $\bf r$.
%
%

Hence, in the non-collinear case, the Hamiltonian cannot be spin-diagonalised by a rotation in spin-space and
for many-electron systems the mapping 
between spin-potentials and ground states is invertible \cite{path}.

Approximate collinear functionals (in the Local-Spin-Density Approximation, LSDA and 
Generalized-Gradient Approximation, GGA) have been employed to 
%
simulate the non-collinear exchange and correlation 
functional $E_{xc}$, 
\cite{kubler,sandratskii,kleinman}: 
The magnetization density is rotated to locally spin-diagonal form: 
${\bf m}({\bf r}) \cdot \mbox{\boldmath $\sigma$} 
\rightarrow $ $M_z({\bf r}) \, \sigma_z $ $= {\sf U} ({\bf r}) \, {\bf m}({\bf r}) 
\cdot \mbox{\boldmath $\sigma$} \, {\sf U}^{\dag} ({\bf r}) $, 
then, the locally diagonal spin-magnetization $M_z({\bf r})$, 
is viewed momentarily as if it were the magnetization density of a 
collinear system. 
The corresponding (exchange and correlation) collinear magnetic field $B_z({\bf r})$ is found by taking 
the functional derivative of the collinear exchange and correlation functional $E_{xc}^{0}$: 
$B_z({\bf r}) = \delta E_{xc}^{0}[\rho , M_z] / \delta M_z({\bf r})$. 
Then, $B_z({\bf r})$ (seen as the magnetic field that 
would have resulted by properly taking the functional derivative 
$\delta E_{xc}[\rho , M_z] / \delta M_z({\bf r})$ 
of the non-collinear 
functional)  
%
is rotated back, 
$B_z({\bf r}) \, \sigma_z \rightarrow $
${\bf B}({\bf r}) \cdot \mbox{\boldmath $\sigma$} 
$  $= {\sf U}^{\dagger } ({\bf r}) \, B_z({\bf r}) \, \sigma_z \, {\sf U} ({\bf r}) $, 
to obtain the non-collinear 
magnetic field ${\bf B}({\bf r})$. 
Finally Pauli-like KS equations are solved using ${\bf B}({\bf r})$. 
%
 Perturbative corrections (spin-stiffness) 
for the non-collinear exchange energy have been proposed \cite{kleinman}. 
The method has given access, successfully, to a wide range of systems 
exhibiting non-collinear magnetism \cite{kubler,sandratskii}, 
notably the spiral spin-density-wave gs of $\gamma$-Fe \cite{bylander,knopfle}.

Next, we analyse the consequences of using the collinear rather than the 
non-collinear exchange and correlation energy for the functional
derivative. 
The difference is the absence of the vector potential 
in $E_{xc}^{0}[\rho , M_z]$. 
We shall investigate the effect this omission for 
weak $\mbox{\boldmath $\sf A$}$.

Following Korenman, Murray and Prange \cite{korenman}, 
the transformed kinetic energy operator (\ref{Tgauge}) can be expressed as a sum of three terms, 
$ 
{\hat T} = {\hat T}_0 + {\hat T}_{ {\sf A}} + {\hat V}_{ {\sf A}}
$, where 
${\hat T}_0$ has the familiar form (\ref{T}) in terms of the rotated field operators (\ref{rot1},\ref{rot2}). 
${\hat T}_{ {\sf A}}$ is linear in $\mbox{\boldmath $\sf A$}$ and describes the interaction of $\mbox{\boldmath $\sf A$}$ with  spin-currents, 
${\hat T}_{ {\sf A}}   =  {\hat T}_{ {\sf A} x} +  {\hat T}_{ {\sf A} y} +  {\hat T}_{ {\sf A} z} $, with
\begin{equation}
{\hat T}_{ {\sf A} \alpha }   =  \hbar  \int d {\bf r} \, {\hat {\bf J}}_{\alpha } ({\bf r}) 
\cdot {\bf A}_{\alpha } ({\bf r})  \, , \ \alpha = x,y,z
\end{equation}
The spin-current operators ${\hat {\bf J}}_{\alpha } ({\bf r})$ are given by (\ref{j}) after 
replacing the unrotated field operators by the rotated ones (\ref{rot1},\ref{rot2}) and 
${\bf A}_{ x (y)}  =  \sin{(\theta + \phi)} \mbox{\boldmath $\nabla$} \omega  - (+)  \, \cos{(\theta + \phi)} \, \sin{\omega} $ $\cos{\omega}$ \mbox{{\boldmath $\nabla$}$ (\theta - \phi)  $}, 
${\bf A}_z = -  \cos^2{\omega} \, \mbox{\boldmath $\nabla$} \theta 
-  \sin^2{\omega} \, \mbox{\boldmath $\nabla$} \phi $. \\ 
${\hat V}_{ {\sf A}}$ contains the square of $\mbox{\boldmath $\sf A$}$ and has the form of a 
potential energy operator:
\begin{equation} \label{t3}
{\hat V}_{ {\sf A}} = \int d{\bf r}
\, {\hat \rho} ({\bf r}) \, V_{ {\sf A}} ({\bf r}) 
\end{equation}
where, $V_{ {\sf A}} ({\bf r}) = (\hbar^2/ 2 m) \, \left( A^2_x 
({\bf r}) + A^2_y ({\bf r}) + A^2_z ({\bf r}) \right) $ and 
${\hat \rho} ({\bf r})$ is the 
density operator $\sum_{\tau } 
{\hat \psi}^{\dag}_{\tau }({\bf r}) {\hat \psi}_{\tau }({\bf r})$. 


The non-interacting kinetic energy functional $T_s[\rho, {\bf m}] \doteq 
\min_{\Phi \rightarrow \rho, {\bf m}} \langle \Phi | {\hat T} | \Phi \rangle $
is written in rotated spin-space, 
$ T_s[\rho, {\bf m}] =  \langle \Phi_{\rho, {\bf m}} | {\hat T}_0 + {\hat T}_{\sf A} 
+ {\hat V}_{\sf A}| \Phi_{\rho, {\bf m}} \rangle \, , 
$
where $\Phi_{\rho, {\bf m}}$ is the KS state (imp-s).    

We note $M_z({\bf r})$ represents (in rotated spin-space) the same magnetization density as 
${\bf m}({\bf r})$.  
We further define:
\begin{equation}
T_s^{0 }[\rho, M_z]  \doteq   \min_{\Phi \rightarrow \rho, M_z} \langle \Phi | {\hat T}_0 | \Phi \rangle  
=  \langle \Phi_{\rho, M_z} | {\hat T}_0 | \Phi_{\rho, M_z} \rangle  
\end{equation}
$T_s^{0}[\rho, M_z]$ would be the collinear non-interacting kinetic energy functional if 
$M_z$ were a collinear magnetization density in the laboratory spin-space. 
The minimizing (KS) state $\Phi_{\rho, M_z}$ is p-s.
For weak $\sf A$ we can approximate $T_s[\rho, {\bf m}]$ by $T_s^{0}[\rho, M_z]$ plus a correction:
\begin{eqnarray} 
\lefteqn{
T_s[\rho, {\bf m}] = T_s^{0}[\rho, M_z] } \label{Tsexpand} \\
&& + \langle \Phi_{\rho, M_z} | {\hat T}_{{\sf A}_z} | \Phi_{\rho, M_z} \rangle 
+  \int d{\bf r} \, \rho({\bf r}) \, V_{\sf A} ({\bf r})  + \Delta T_s[\rho, M_z]  \nonumber
\end{eqnarray}

A similar analysis can be carried out for the internal energy functional, 
$
F[\rho, {\bf m}] = \min_{\Psi \rightarrow \rho, {\bf m}} \langle \Psi | {\hat T} + {\hat V}^{\rm ee}| 
\Psi \rangle  
$. 
Replacing $\hat T$ by ${\hat T} + {\hat V}^{\rm ee}$ above, 
we obtain $F^{0}[\rho, M_z]$, $\Delta F[\rho, M_z]$ in place of 
$T_s^{0}[\rho, M_z]$, $\Delta T_s[\rho, M_z]$ and 
the minimizing states $\Psi_{\rho, {\bf m}}, \Psi_{\rho, M_z}$ in place of 
$\Phi_{\rho, {\bf m}}, \Phi_{\rho, M_z}$.

The non-collinear exchange and correlation functional, 
$ 
E_{xc}[\rho, {\bf m}]$ $ \doteq$ $ F[\rho, {\bf m}] - T_s[\rho, {\bf m}] - (1/2) \int 
\!\! \rho({\bf r}) \rho({\bf r}') / |{\bf r} - {\bf r}'| 
$,
becomes  
\begin{eqnarray} \label{Excexpand}
E_{xc}[\rho, {\bf m}]  & = &    E_{xc}^{0}[\rho, M_z] + 
\langle \Psi_{\rho, M_z} | T_{{\sf A}_z} | \Psi_{\rho, M_z} \rangle \\
&&  
- \langle \Phi_{\rho, M_z} | T_{{\sf A}_z} | \Phi_{\rho, M_z} \rangle + \Delta E_{xc}[\rho, M_z]  \nonumber
\end{eqnarray}
where, $\Delta E_{xc}[\rho, M_z]  = \Delta F[\rho, M_z]   - \Delta T_{s}[\rho, M_z]$ and
$
E_{xc}^{0}[\rho, M_z] = F^{0}[\rho, M_z] - T_{s}^{0}[\rho, M_z] - (1/2) \!\! \int 
\rho({\bf r}) \rho({\bf r}') / |{\bf r} - {\bf r}'|$. 
When $M_z$ represents a collinear magnetization density in the laboratory spin-space, then 
$E_{xc}^0[\rho, M_z]$ is the collinear exchange and correlation energy functional. 

In the non-collinear LSDA/GGA method \cite{kubler,sandratskii}, instead of the non-collinear 
$E_{xc}[\rho, {\bf m}]$, one deals with $E_{xc}^{0}[\rho, M_z]$.
From Eq. \ref{Excexpand}, there are corrections, $\langle \Psi_{\rho, M_z} | 
T_{{\sf A}_z} | \Psi_{\rho, M_z} \rangle - 
\langle \Phi_{\rho, M_z} | T_{{\sf A}_z} | \Phi_{\rho, M_z} \rangle$
(first-oder) and $\Delta E_{xc}[\rho, M_z]$ (second and higher).
These corrections are given in terms of p-s states $\Phi_{\rho, M_z}$, 
$\Psi_{\rho, M_z}$ and it is possible to approximate them perturbatively, using 
the spin-polarized uniform electron gas model, in order to 
obtain a fully non-collinear LSDA exchange and correlation functional.   

Recently, the formalism was developed to treat exchange exactly for non-collinear magnets \cite{helbig}.
Both approaches have advantages (simplicity of calculation, access to larger systems, 
some account of correlation for non-collinear LSDA/GGA, compared with accurate 
treatment of non-collinearity and elimination of self-interaction for exact-exchange). 

In conclusion, we have analysed the non-collinear exchange and correlation 
energy functional. 
It is intriguing to derive and observe the effect of corrections we have suggested 
and to compare results with the exact-exchange scheme. 
We remark that any KS approach, crucially relies on the unique determination of the KS potential 
by the KS state that this study ensures. 

I am grateful to K. Capelle, H. Eschrig, W. Kohn, J. K\"ubler, W.E. Pickett for useful comments. 
I thank M. Levy for discussions and encouragement at an early stage.


\begin{thebibliography}{99}
\bibitem{path}
Barring pathological cases, where the ground state of the spin-diagonal part of $\mathcal H$ in (\ref{one}) is also an eigenstate of its non-diagonal part, even though the two operators do not commute. 
See Footnote 18 in \cite{argaman}.
\end{thebibliography}
\end{document}